\documentclass[%
	final,
	notitlepage,
	narroweqnarray,
	inline,
	twoside,
	]{ieee}

\begin{document}

\title[]{Deriving Specifications of \\ Dependable Systems: \\ toward a Method}

\author[]{Manuel Mazzara \\ School of Computing Science, Newcastle University, UK \\ manuel.mazzara@newcastle.ac.uk}

\maketitle               

\begin{abstract} 
This paper proposes a method for deriving formal specifications of systems. To accomplish this task we pass through a non trivial number of steps, concepts and tools where the first one, the most important, is the concept of method itself, since we realized that computer science has a proliferation of languages but very few methods. We also propose the idea of Layered Fault Tolerant Specification (LFTS) to make the method extensible to Dependable Systems. The principle is layering the specification, for the sake of clarity, in (at least) two different levels, the first one for the \textit{normal behavior} and the others (if more than one) for the \textit{abnormal}. The abnormal behavior is described in terms of an Error Injector (EI) which represents a model of the erroneous interference coming from the environment. This structure has been inspired by the notion of idealized fault tolerant component but the combination of LFTS and EI using rely guarantee thinking to describe interference can be considered one of the main contributions of this work. The progress toward this method and the way to layer specifications has been made experimenting on the Transportation and the Automotive Case Studies of the DEPLOY project.
\end{abstract}

\section{Introduction}

\begin{quote}
\textit{"Dubium Sapientiae initium" - Descartes}
\end{quote}

There is a long tradition of approaching Requirements Engineering (RE) by means of formal or semi-formal techniques. Although "fuzzy" human skills are involved in the process of elicitation, analysis and specification - as in  any other human field - still methodology and formalisms can play an important role \cite{SMART}. The first thing we realized in building dependable software is the necessity to build dependable communication between parties that use different languages and vocabulary. In order for the systems to match expectations (and specifications), we need indeed a precise mapping between intentions and actions. Formal methods appear to be an effective solution.

Object Oriented Design \cite{BoochOO} and Component Computing \cite{SzyperskiCS} are just well known examples of how some rigor and discipline can improve the final quality of software artifacts besides the human communication factor. The success of languages like Java or C\# could be interpreted in this sense, as natural target languages for this way of structuring thinking and design. It is also true - and it is worth reminding it - that in many cases it has been the language and the available tools on the market that forced designers to adopt object orientation principles, for example, and not vice versa. This is the clear confirmation that it is always a combination of conceptual and software tools together that create the right environment for the success of a discipline.

Semi formal notations like UML \cite{FowlerUML} helped in creating a language that can be understood by both specialists and non specialists, providing different views of the system that can be negotiated between different stakeholders with different backgrounds. The power (and thus the limitation of UML) is the absence of a formal semantics (many attempts can be found in the literature anyway) and the strong commitment on a way of reasoning and structuring problems which is clearly the one disciplined by object orientation. Many other formal/mathematical notations existed for a long time for specifying and verifying systems like process algebras (a short history by Jos Baeten in \cite{BaetenPA}) or specification languages like Z (early description in \cite{AbrialZ-book}) and B \cite{AbrialB-book}. The Vienna Development Method (VDM) is maybe one of the first attempts to establish a Formal Method for the development of computer systems \cite{BjornerVDM78}. All these notations are very specific and can be understood only by specialists. The point about all these formalisms is that they are indeed notations, formal or semi-formal. Behind each of them there is a way of structuring thinking that does not offer complete freedom and thus forces designers to adhere to some discipline. But still they are not methods in the proper sense, they are indeed languages. 

The goal of this paper is providing a different view for interpreting problems and faults. The overall result will be the definition of a method for the specification of systems that do not run in isolation but in the real, physical world. To accomplish this task we need to pass through a non trivial number of steps, concepts and tools where the first one, the most important, is the concept of method itself, since we realized that computer science has a proliferation of languages but very few methods. In the following we want to put more emphasis on the difference between methods and languages and, as a consequence, between formal methods and formal languages.

\section{About Methods and Languages}

The idea of this section is defining a set of desiderata for the method we will present later in this paper. We reached these ideas partly in an attempt to understand what a method basically is, and partly gaining experience and insights by experimenting with the DEPLOY pilot projects \cite{DEPLOY}. Firstly, we think it is important to distinguish between the words method and methodology, often misused. For the Webster's dictionary "a method is a way, technique, or process of or for doing something". It is worth noting that this definition depends on the one of process which is "a series of actions or operations conducing to an end". The word methodology can also be used to intend "a particular procedure" but the general meaning is "the analysis of the principles of methods [...] employed by a discipline". According to these definitions, when we refer to Process Algebras, for example, the words methodology and method are not correct. Although in computer science it is common practice to use the word formal methods to intend formal languages, in this paper we will use the word method only to intend the final result of a methodological study related to a specific context, software systems specification  in this case. 

To properly understand what a method is and what it is not we explored an illustrious example by Descartes, the "Discourse on the Method" (1637) \cite{Descartes}. It is the famous philosophical and mathematical treatise which is the source of the quotation "Je pense, donc je suis" ("I think, therefore I am"). For lack of space here it is not possible to report all the relevant parts. We are only interested in understanding how Descartes perceives a method and what is peculiar in it. Looking at \cite{Descartes} it is easy to realize that a method proposes a partially ordered set of actions that need to be performed and then discharged within a specific causal relationship. The success of one action determines the following ones. More in detail, this study suggested us that the method we are going to propose needs to satisfy a number of properties. The first three are taken directly from Descartes' method, while the last three are grabbed from our experiments with case studies which we will partly introduce later but we think it is worth to gather all the features together now. Our understanding of the method of science has told us that:
\\
\begin{enumerate}
	\item It has to consist of steps to acquire knowledge
	\item It has to be formally defined ("phases", steps, workflow…)
	\item It has to be repeatable (by non formal methods experts)\\
\end{enumerate}

Then our practical experience has suggested that:
\\
\begin{enumerate}
	\item It has to be scalable (non "ad hoc" - it has to work outside specific case studies)
	\item It needs abstractions (what and not how)
	\item It has to be extensible to fault tolerant behavior (we propose the idea of LFTS for this)\\
\end{enumerate}

The above definitions and the "Discourse on the Method" are a starting point to understand better what a method is and what it is not. At this point the differences between a formal language and a formal method should be clearer. Now we have to ask ourselves why we need a method. The "why" is an interesting point, it is a meta question, a question that allows us to reason about "the method" looking from outside the method. The logic is what is done "inside the system", in this case the formal steps performed (in some order) to reach the desired end i.e. the method itself. The reasoning is what is done "outside the system", experimenting and seeing what happens if we change the basic rules. Reasoning "about the method" gives us a way to find out the motivations leading to a method definition. What we believe now is that the first step in building dependable software is building dependable communication between parties that have different languages/vocabulary. According to the definition of communication \cite{DeFleur93}, formal methods in system specification are tools to commit on dependability since they help us in clarifying our vocabulary and providing a notation able to build a precise mapping between intentions and actions in the different stakeholders' minds. Thus a clear and precise definition of a formal method (in the actual meaning of the word) seems to be necessary at this stage. 

After having understood where we want to go and why, now it is good practice to say how we want to get there.

\section{Toward a Method}

Our work in this paper focuses especially on \cite{JonesHJ07} where the original idea of a formal method for the specification of systems running in the physical world originated. That paper was full of interesting ideas but still was lacking of a method in the sense we described so far. Few case studies have been analyzed according to this philosophy in \cite{Jones05t} but still a complete method has not been reached. For this reason now we think that a more structured approach is urgent in this area. Thus, the goal of the present work is to improve our understanding of those ideas, trying to increment that contribution and to put it in an homogeneous and uniform way describing a method featuring the properties we introduced above with particular attention to dependable systems. At the moment we have had some progress in this direction but we still need more work. The basic idea behind \cite{JonesHJ07} was to specify a system not in isolation but considering the environment in which it is going to run and deriving the final specification from a wider system where assumptions have been understood and formalized as layers of rely conditions (we will explain this later in the paper). Here the difference between assumptions and requirements is crucial, especially when considering the proper fault tolerance aspects. We could briefly summarize this philosophy as follows:
\\
\begin{itemize}
\item Not specifying the digital system in isolation
\item Deriving the specification starting from a wider system in which physical phenomena are measurable
\item Assumptions about the physical components can be recorded as layers of rely-conditions (starting with stronger assumptions and then weakening when faults are considered)\\
\end{itemize}

This approach allows us to see a computer system from a different angle, as not consisting of functions performing tasks in isolation but as relationships (interfaces/contracts) in a wider world including both the machine and the physical (measurable) reality. This philosophy has been partly inspired by Michael Jackson's approach to software requirements analysis typically called Problem Frames approach \cite{JacksonPF}. The software running on the hosting machine is called Silicon Package and it should be clear that the machine itself can neither acquire information on the reality around nor modify it. The machine can only operate trough sensors and actuators. To better understand this point, we like to use a similar metaphor about humans where it is easier to realize that our brain/mind system (our Silicon Package?) cannot acquire information about the world  but it can only do that through eyes, ears and so on (our sensors). In the same way it cannot modify the world if not through our arms, voice, etc (our actuators). So, as we start describing problems in the real world in terms of what we perceive and what we do (and not about our brain functioning) it makes sense to adopt a similar philosophy for computer systems consisting of sensors and actuators. Around the Silicon Package we have the problem world and the assumptions that need to be made regarding it. We want to derive the specification of the Silicon Package starting from the wider system. The way in which we record these assumptions is a topic for the following sections. 

\subsection{The Method and its Steps}

In this work we are structuring the method introduced in \cite{JonesHJ07} according to the properties described in the previous section. To do this behind that work we recognize three main steps:
\\
\begin{enumerate}
	\item Define boundaries of the systems
	\item Expose and record assumptions (by means of rely-conditions)
	\item Derive the specification\\
\end{enumerate}

Our idea is to not commit to a single language/notation - we want a formal method, not a formal language - so we will define a general high level approach following these guidelines and we will suggest, in the case study, \textit{reference tools and notations} to cope with these steps. It is important to note that these are only reference tools that are \textit{suggested} to the designers because of a wider experience regarding them from our side. A formal notation can be the final product of the method but it still needs to be not confused with the method itself. In this work we want to emphasize the different steps that were not clearly defined previously in \cite{JonesHJ07}. The reader will understand that this is still a simplification of the process. We use the word "steps" instead of "phases" since we do not want to suggest a sort of linear process which is not always applicable, in the average case (especially when coping with fault tolerance as we will discuss later). We imagine, in the general case, many iterations between the different steps. The idea of the method is to ground the view of the Silicon Package in the external physical world. This is the problem world where assumptions about the physical components \textit{outside} the computer itself have to be recorded. Only after this can we derive the specification for the software that will run \textit{inside} the computer. This more precise formalization of the method and the features the method has to exhibit is one of the main contributions of this work. 

\subsection{Rely/Guarantee Thinking and Interference}

We have already mentioned the Rely/Guarantee thinking a number of times in this paper but to really understand its power it is necessary to realize how preconditions and postconditions can help in specifying a software program when interference does not play a role. What we have to describe (by means of logical formulas) when following this approach is:
\\
\begin{enumerate}
	\item the input domain and the output range of the program
	\item the precondition, i.e. the predicate that we expect to be true at the beginning of the execution
	\item the postcondition, i.e. the predicate that will be true at the end of the execution provided that the precondition holds\\
\end{enumerate}

Preconditions and postconditions represent a sort of contracts between parties: provided that you (the environment, the user, another system) can ensure the validity of a certain condition, the implementation will modify the state in such a way that another known condition holds. We show the example of a very simple program, the specification of which in the natural language may be:\\

\textit{"Find the smallest element in a set of natural numbers"}\\\\
This very simple natural language sentence tells us that the smallest element has to be found in \textit{a set of natural numbers}. So the output of our program has necessarily to be a natural number. The input domain and the output range of the program are then easy to describe:

\small

$$I/O:  \mathcal{P}(\textbf{N}) \rightarrow \textbf{N}$$ 

\normalsize

\noindent
Now, the input is expected to be a set of natural numbers, but in order to be able to compute the min such a set has to be non empty since the min is not defined for empty sets. So the preconditions that has to hold will be (please, note the difference between $\mathcal{P}$ and P):

\small

$$P(S):  S \neq \emptyset$$

\normalsize

\noindent
Provided that the input is a set of natural numbers \textit{and} it is not empty, the implementation will be able to compute the min element which is the one satisfying the following:

\small

$$Q(S,r): r \in S  \wedge (\forall e \in S)(r \leq e) $$

\normalsize

\noindent
Given this set of rules, the input-output relation is given by the following predicate that needs to be satisfied by any implementation \textit{f}:

\small

$$\forall S \in \mathcal{P}(\textbf{N}) (P(S) \Rightarrow f(S) \in  \textbf{N} \wedge Q(S, f(S)))$$ 

\normalsize

Now, to better understand the limitations of these kind of abstractions, consider the case of interference happening through a global state. Two processes alternate their execution and access to the state. The global state can consist of shared variables or can be a queue of messages if message passing is the paradigm adopted (at the end the two paradigms are equivalent). This is exactly the situation described in \cite{Jones83a}. In the case in which we consider interfering processes we need to accept that the environment can alter the global state but the idea behind R/G is that we impose these changes to be constrained. Any state change made by the environment (other concurrent processes with respect to the one we are considering) can be assumed to satisfy a condition called R (rely) and the process under analysis can change its state only in such a way that observations by other processes will consist of pairs of states satisfying a condition G (guarantee). Thus, the process \textit{relying} on the fact that a given condition holds can \textit{guarantee} another specific condition. Consider, for example, the two following simple pieces of code, the cooperation of which calculates the Greatest Common Divisor:

\small
$$
\begin{array}{lll}
\texttt{P1:} 					               & \; \; \; \; \; \; \; \; \; \; \; \; \; \; \; \; \; \; \; \; \; \; \; \; & \texttt{P2:} \\
\texttt{while(a<>b)} \texttt{\{}     & \; \; \; \; \; \; \; \; \; \; \; \; \; \; \; \; \; \; \; \; \; \; \; \; & \texttt{while(a<>b)\{} \\
\; \; \; \;  \texttt{if(a > b)}      & \; \; \; \; \; \; \; \; \; \; \; \; \; \; \; \; \; \; \; \; \; \; \; \; & \; \; \; \; \texttt{if(b > a)} \\
\; \; \; \; \; \; \texttt{a := a-b;} & \; \; \; \; \; \; \; \; \; \; \; \; \; \; \; \; \; \; \; \; \; \; \; \; & \; \; \; \; \; \;  \texttt{ b := b-a;}  \\
\texttt{\}} 					               & \; \; \; \; \; \; \; \; \; \; \; \; \; \; \; \; \; \; \; \; \; \; \; \; & \texttt{\}}				 
\end{array}
$$

\normalsize
\noindent
P1 is in charge of decrementing \textit{a} and P2 of decrementing \textit{b}. When $a=b$ will evaluate to true it means that one is the Greatest Common Divisor for \textit{a} and \textit{b}. The specification of the interactions is as follows:\\

\noindent
\small
$R_1:(a=\overline{a}) \wedge (a \geq b \Rightarrow b= \overline{b}) \wedge (GCD(a,b) = GCD(\overline{a},\overline{b}) )$\\
$G_1:(b=\overline{b}) \wedge (a \leq b \Rightarrow a=\overline{a}) \wedge (GCD(a,b) = GCD(\overline{a},\overline{b}) ) $\\
$R_2 = G_1$\\
$G_2 = R_1$\\

\normalsize
\noindent
Here the values $\overline{a}$ and $\overline{b}$ are used instead of \textit{a} and \textit{b} when we want to distinguish between the values before the execution and the values after. P1 relies on the fact that P2 is not changing the value of \textit{a} and $a \geq b$ means no decrements for  \textit{b} have been performed. Furthermore the GCD did not change. Specular situation is for the guarantee condition. Obviously, what is a guarantee for P1 becomes a rely for P2 and vice versa.

\section{The Transportation Case Study}

The ideas presented in this paper about the method and the way to organize specifications derived by experimenting with interesting case studies through the development of the DEPLOY project \cite{DEPLOY}. We have mainly got experience from two studies: namely the Transportation and the Automotive Case Studies (DEPLOY WP1 and WP2). In this work we describe the first one of these studies and in particular we will show how it brought us to a better definition of the method. The case study is taken from \cite{AbrialEvent-B}, the train system, where the goal was showing the power of modeling and formal reasoning by means of Event-B examples. We chose this scenario since we believe it is particularly realistic (it has been developed after some work with real train systems) and still manageable (with a limited set of initial requirements: 39). This case study taught us how to distinguish between assumptions and requirements and helped us in finding a better structure for the method initially presented in \cite{JonesHJ07}. We will show here how this example can be approached with the three step method. The first thing to do is deciding the bounds of specification (step 1). We will then show how the boundaries can be broadened to include the external world. In the second step we will discuss how to separate assumptions and requirements, how to expose and record assumptions and how different sets of requirements and assumptions will imply a different specification and then implementation. In the third step we will assume the existence of an already designed network infrastructure (with sensors etc...) to show a specific example of implementation. At the end we will show how to make use of rely conditions for this specific implementation. Unfortunately, because of space constraints, it is not possible to present all the requirements and the necessary background (the interested reader can find all the details in \cite{AbrialEvent-B}). What it is important to realize is the way in which the interference over a global state is considered using the approach showed in the small GCD example. In the following, the specification will be showed, after a discussion about the way in which it has been obtained, and the interaction between the different operations will be constrained in a similar way but in a system with a potentially higher level of concurrency.

\subsection*{Step 1: defining the boundaries of the system}

The idea here, as said, is to clarify the requirements in the real world before trying to specify the software which sits within the system. This process naturally identifies assumptions about the physical components which can be then recorded as rely-conditions. One of the main principles of this approach is not specifying a system in isolation but starting to move the system boundaries outwards (what is called "pushing out the boundaries of the system" in \cite{JonesHJ07}). What is the wider system in which physical phenomena are measurable? What is the actual general purpose of the Train System? We believe it is allowing trains to move safely from a place X to a place Y. How does this help us in identifying the system requirements?  We can recognize that the FUN-1 requirement of the system specification \cite{AbrialEvent-B} expresses basically this need:
\\
\begin{quote}
"The goal of the train system is to safely \\ control trains moving on a track network"\\
\end{quote}
\noindent
If we move the boundary outwards further we can say that the purpose of the system is allowing people to reach their destination "safely". Considering this we could split FUN-1 in two properties (without referring to any specific implementation yet):
\\
\begin{itemize}
\item Safety property: nothing bad can happen 
\item Liveness property: something good has to happen\\
\end{itemize}
We can express these two properties more in detail for this example as follows:
\\
\begin{itemize}
\item Safety: Trains will never collide
\item Liveness: Trains will move from their origin to their destination\\
\end{itemize}

Req FUN-1 is general enough to allow this separation, anyway we are interested here in modeling only the safety property delegating liveness to a scheduler or, theoretically, to a manual management performed by operators/engineers. For the sake of simplicity, the specification will start with this requirement only. All the other requirements presented in \cite{AbrialEvent-B} refer to concepts like blocks, routes and signals that can describe either a set of assumptions about the environment or a specific implementation of FUN-1. We will say more about this later.

\subsection*{Step 2: exposing and recording assumptions}

Now it is crucial to discriminate between \textit{requirements for the system} and \textit{assumptions about the "real world"}. In this example it was important to ask if we are in charge of designing the whole railway/track with sensors, signals, etc. or not. If not, many of the requirements (and the given block structure showed later) can be considered as assumptions taken from the already existing environment (for example the ENV group of requirements in \cite{AbrialEvent-B}). Otherwise, they can be still seen as requirements but referring to a specific implementation and they should not be introduced now but only later, in the last step. For example, the requirement ENV-13:
\\
\begin{quote}
"A signal can be red or green. Trains \\ are supposed to stop at red signals"\\
\end{quote}
\noindent
is an example of how requirements and assumptions can be (in our opinion erroneously) mixed in the same statement. So determining the assumption (and being able to separate them from requirements) is the main goal of this step. In this example we suppose to be the designer of the whole track and we want trains to move from city X to city Y. There are many possible implementations for the presented requirements, we will look into the details of only one (which is the one given in \cite{AbrialEvent-B}). Before looking at that it is easy to understand that the simplest possible implementation is the one requiring that no train can cross the network. This is an implementation where the Safety property is preserved (but Liveness is not). Although we are interested mainly in this property here, a better thing to do would be allowing only one train on the track between X and Y. This means basically that the rail connecting two cities will be reserved for a single train. Obviously this implementation respects both the safety and liveness requirements described above. But it is also easy to realize that it is simply unfeasible because of the very low efficiency, very low exploitation of the available resources and because of the expensiveness (time and money).

\begin{figure}[htp]
\centering
\includegraphics[scale=0.2]{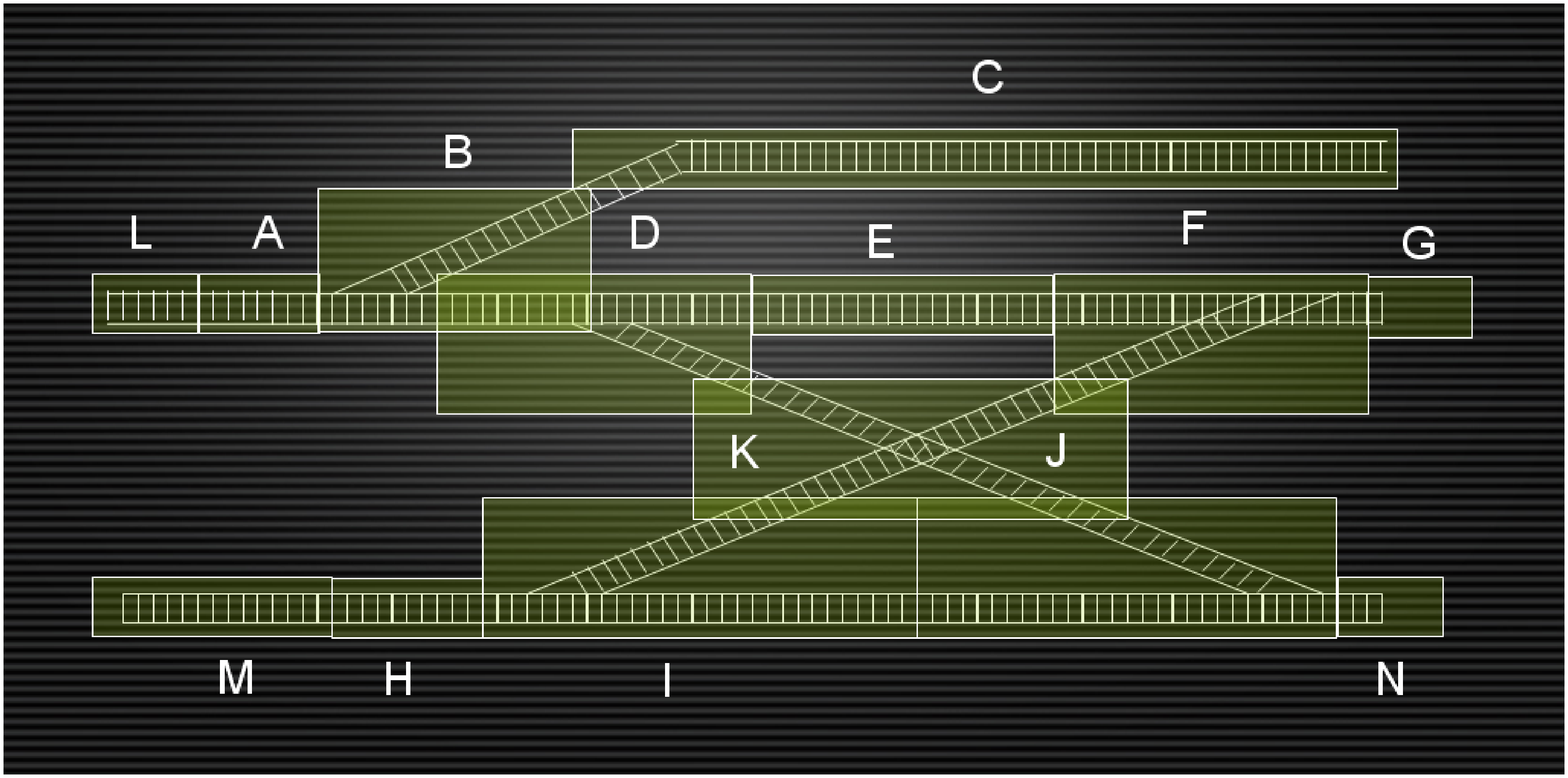}
\caption{The network infrastructure}\label{blocks}
\end{figure}

A more reasonable implementation is actually the one that in \cite{AbrialEvent-B} is simply used for the modeling purposes. The scope here is different from what has been done there, for this reason we did not assume this implementation as given but we wanted to go through the entire discussion. The point was learning the lesson about determining wider boundaries, including the external environment, and distinguishing between requirements and assumptions. So we analyzed the entire process carefully. Now we are ready to present this implementation. Figure \ref{blocks} represents an example of the infrastructure. It is made of:
\\
\begin{enumerate}
\item Blocks: a track is made of a number of fixed blocks as showed in figure \ref{blocks}
\item Routes: blocks are always structured in a number of statically predefined routes. Each route represents a possible path that a train may follow. Routes define the various ways a train can cross the network. A route is composed of a number of adjacent blocks forming an ordered sequence. For example a route is LABDKJN.
\item Points: a track contains these special components. A point may have two positions: directed or diverted. These components are attached to a given block. And a block contains at most one special component. In figure \ref{blocks}, B and D, for example, they both contain points.
\item Signals: each route is protected by a signal (Red/Green). It is situated just before the first block of each route and it must be clearly visible by train drivers.
When a signal is red the corresponding route cannot be used by an incoming train.\\
\end{enumerate}

The idea is to have each block of a reserved route freed as soon as the train does not occupy it anymore. It is not scope of this work to describe entirely the case study but only to describe the insights we have gained in the process of working toward a method. The reader who can find it hard to abstract over few details should refer to \cite{AbrialEvent-B} for the detailed description of this scenario. In the next section we will focus on the reserving routes system, i.e. the process of reserving a route on a train request, freeing it and letting the train occupy block by block freeing each block when passed. 

We have also decided to abstract over concepts like time or distances since the underlying block-based infrastructure will ensure that we will never have two trains in the same block in such a way to avoid collisions. This abstraction simplifies our work without being in contradiction with the original philosophy of grounding the system in the physical world. We have only decided that the border between the system and the real world here will consist of the sensors and actuators necessary to make such an infrastructure working. We have also decided to focus on safety purposes. We indeed do not cope with liveness, we assume that these kind of problems are managed by a scheduler which is another system already running. A graph structure would be probably more adequate in case we want to focus on the scheduler having Liveness coming into play since in the present representation the network is seen as a set of routes from which you cannot really infer which one is adjacent to the other. 
 
\subsection*{Step 3: deriving the formal specification}
Now we define the basic machinery for the formal specification. We need four finite sets for the purpose:\\
\begin{itemize}
	\item \texttt{T}, a finite set of trains (\textit{t} a variable ranging over it)
	\item \texttt{B}, a finite set of blocks (\textit{b} a variable ranging over it)
	\item \texttt{R}, a finite set of routes (\textit{r} a variable ranging over it)
	\item \texttt{P}, a finite set of points (\textit{p} a variable ranging over it)\\
\end{itemize}

The safety requirement will be modeled as a total function mapping blocks to trains: $\texttt{B} \rightarrow \texttt{T}\;(\texttt{train})$. This is how we impose to have a single train on a block. To avoid collisions by trains we also need a way to associate trains to routes, once the train has reserved a specific route. We use the function: $\texttt{T} \rightarrow \texttt{R} \;(\texttt{route})$. A route is then composed by blocks, at least one: $ \texttt{R} \rightarrow \texttt{B$^{+}$} \;(\texttt{blocks})$ and in a route a block has a next element: $ \texttt{B} \rightarrow \texttt{B} \;(\texttt{next})$. Blocks can be free or occupied : $ \texttt{B} \rightarrow \{free,occupied\} \;(\texttt{status})$ and are associated to points: $ \texttt{B} \rightarrow \texttt{P} \; (\texttt{point})$ that can be oriented in two different ways: $ \texttt{P} \rightarrow \{directed,diverted\} \;(\texttt{direction})$. Routes can be available or reserved: $ \texttt{R} \rightarrow \{available,reserved\} \;(\texttt{availability})$ and each route is associated with a predefined points orientation: $ \texttt{R} \rightarrow (\texttt{P} \rightarrow \{directed,diverted\}) \;(\texttt{orientation})$. We rely on the fact that the sensors with which a block is equipped can always detect the presence of a train (for $\texttt{B} \rightarrow \texttt{T}$). We assume that if we want to reserve a point, it will be promptly positioned. We do not model these "low level" aspects here (for $\texttt{T} \rightarrow \texttt{R}$). We rely also on the fact that each route has a first block: $\texttt{R}\rightarrow \texttt{B}$ (\textbf{first}), a last block: $\texttt{R}\rightarrow \texttt{B}$(\textbf{last}), and they are different: $first(R) \neq last(R)$. 

The mathematical machinery defined so far can be considered part of the global state on which the five operations we are going to define operate: they are related to the process of route reservation and freeing plus the entrance, proceeding and exit of a train to and from a route. These are the operations concerned with the specification of our safety requirement. Liveness is not discussed, we only move a train from one end of a route to the other without investigation about the way in which the routes are previously organized. For each operation the notation below indicates the data needed and what we expect from that data plus the way in which the global state will be modified.  

\small
$$ 
\begin{array}{ll}
\textbf{Operation} 	&	RouteReserving \;(\textit{t}:\texttt{T}, \textit{r}:\texttt{R}) \\
\textbf{Rely}   		& availability(r) = available \\
										& \forall \; b \; \in \; blocks(r) \; (status(b)=free)  \\
\textbf{Guarantee}  & availability(r) := reserved  \\ 
										& \forall \; b \in \; blocks(r) \; (status(b):=occupied) \\
										& route(t):=r \\
										& \forall \; p \in \; P \; (direction(p):=orientation(r)(p))
\end{array}
$$
\normalsize
Given a train and a route, this operation guarantees three mappings to be properly updated, provided that the given route is available and the related blocks are free. The three mappings are first the one between points and directions, second the one between  trains and routes (as a record of the overall track status) and last the association between blocks and their occupancy status. These represent the part of the global state of interest for this operation. 

\small
$$ 
\begin{array}{ll}
\textbf{Operation} 	& RouteFreeing \; (\textit{t}:\texttt{T}) \\
\textbf{Rely}   		& \forall b \in blocks(route(t)) \; (status(b)= free)  \\
\textbf{Guarantee}  & availability(route(t)):=available \\
										& route(t):=null 		 		
\end{array}
$$
\normalsize
Given a train the related route is identified. The effect on the state is a modification of the mapping where the train is associated to the null route and, provided that all the blocks in the route are free, the route itself can be freed. This operation has a simpler definition with respect to the reservation because the blocks are freed by the \textit{ExitRoute} while the points direction does not need to be modified when freeing a route.

\small
$$
\begin{array}{ll}
\textbf{Operation}  & EnterRoute \; (\textit{t}:\texttt{T})   \\
\textbf{Rely}    		& availability(route(t)) = reserved \\
                 		& \wedge \; status(first(route(t))) = free \\
\textbf{Guarantee}  & status(first(route(t))):=occupied				 
\end{array}
$$
\normalsize
This operation corresponds to a train entering the first block of a route. The first block must be unoccupied before the operation and it will be occupied afterward. It can be accessed only by trains that have already reserved a route.

\small
$$
\begin{array}{ll}
\textbf{Operation}    & MovingOnRoute \; (\textit{t}:\texttt{T}, \textit{b}:\texttt{B}) \\
\textbf{Rely}   			& availability(route(t)) = reserved  \\
											& \wedge \; b \in blocks(route(t))\\
                			& \wedge \; status(next(b)) = free \\
\textbf{Guarantee}   	& status(b):=free \\
											& status(next(b)):=occupied			 
\end{array}
$$
\normalsize
This operation corresponds to the occupancy of a block which is different from the first block of a reserved route. It can be accessed only by trains that have already reserved a route. The current block has to belong to the route and the next one can be occupied only when it is free. The occupation of the next block implies that the current one becomes free. 

\small
$$
\begin{array}{ll}
\textbf{Operation}     		& ExitRoute \;(\textit{t}:\texttt{T}, \textit{b}:\texttt{B}) \\
\textbf{Rely}   					& availability(route(t)) = reserved  \\
													& \wedge \; b \in blocks(route(t))\\
													& \wedge \; next(b) = \emptyset \\
\textbf{Guarantee}   			& \forall b \in blocks(route(t)) \; status(b):=free  				 
\end{array}
$$
\normalsize
This operation corresponds to the train exit out of the route. It can be accessed only by trains that have already reserved a route and it is responsible to free all the blocks in that route.

\section{Layered Fault Tolerant Specification}

The previous sections discussed how to derive a specification of a system looking at the physical world in which it is going to run. No mention has been made of fault tolerance and abnormal situations which deviate from the basic specification. The first thing the reader will realize is that the method we defined does not cope with these issues but it does not prevent fault tolerance from playing a role. The three steps simply represent what you have to follow to specify a system and they do not depend on what you are actually specifying. This allows us to introduce more considerations and to apply the idea to a wider class of systems. Usually in the formal specification of sequential programs widening the precondition leads to make a system more robust. The same can be done by weakening rely conditions. For example, if eliminating a precondition the system can still satisfy the requirements this means we are in presence of a more robust system. Here we intend to promote this approach combined with the idea of fault as interference. Quoting \cite{ColletteJ00}:
\\
\begin{quote}
The essence [...] is to argue that faults can be viewed as interference in the same way that concurrent processes bring about changes beyond the control of the process whose specification and design are being considered.\\
\end{quote}

In this work we introduce the idea of Layered Fault Tolerant Specification (LFTS) combining it with the approach quoted above and making use of rely/guarantee thinking. The first step in this direction is defining a Fault Model, i.e. which kind of abnormal situations we are considering. Our specification will then take into account that the software will run in an environment when specific things can behave in an "abnormal" way. There are three main abnormal situations in which we can incur, they can be considered in both the shared variables and message passing paradigm: 
\\
\begin{itemize}
	\item Deleting state update: "lost messages"
	\item Duplicating state update: "duplicated messages"
	\item Additional state update (malicious): "fake messages created" \\
\end{itemize}

The first one means that a message (or the update of a shared variable) has been lost, i.e. its effect will not be taken into account as if it never happened. The second one regards a situation in which a message has been intentionally sent once (or a variable update has been done once) but the actual result is that it has been sent (or performed) twice because of a faulty interference. The last case is the malicious one, i.e. it has to be done intentionally (by a human, it cannot happen only because of hardware, middleware or software malfunctioning). In this case a fake message (or update) is created from scratch containing unwanted information.

In our approach the model of fault is represented by a so called \textit{Error Injector} (EI). The way in which we use the word here is different with respect to other literature where Fault Injector or similar are discussed. Here we only mean a model of the erroneous behavior of the environment. This behavior will be limited depending on the number of abnormal cases we intend to consider and the EI will always play its role respecting the defined R/G rules. The operations will rely on a specific abnormal behavior and, given that, will guarantee the ability to handle these situations. More in detail, the rules are as follows:
\\
\begin{itemize}
	\item The Error Injector (environment) interferes changing the global state but respecting his G  – for example, only "lost messages" can be handled
	\item The operation relying on this kind of (restricted) interference is able to handle exceptional/abnormal (low frequency) situations satisfying a weaker G \\
\end{itemize}

All the possibilities of faults in the system are described in these terms and the specification is organized according to the LFTS principle, i.e layering the specification, for the sake of clarity, in (at least) two different levels, the first one for the \textit{normal behavior} and the others (if more than one) for the \textit{abnormal}. This approach originated from the notion of idealized fault tolerant component \cite{AndersonFT} but the combination of LFTS and rely guarantee reasoning can be considered one of the main contributions of this work. The main motto for LFTS is: "Do not put all in the "normal mode". From the expressiveness point of view, a monolithic specification can include all the aspects, both faulty and non faulty, of a system in the same way as it is not necessary to organize a program in functions, procedures or classes. The matter here is pragmatics, we believe that following the LFTS principles a specification can be more understandable for all the stakeholders involved. We also hope that, considering the Silicon Package in its relationships with the physical world, could assist the process of finding faulty behavior important to the system. 

\subsection*{LFTS for the Train System}

Here we consider the Train System in a less ideal world than the one analyzed before. In this world, the EI plays its role, for the sake of simplicity, changing the global state only according to the "lost messages" condition. The global state of the system needs to be modified for the EI to implement its changes. Now, in the network, sensors and actuators can actually fail and some state update could be not performed. Thus, let us modify the \texttt{availability} function in such a way as to include a third option: $ \texttt{R} \rightarrow \{available,reserved, maintenance!\} \;(\texttt{availability})$. The \textit{RouteReserving} operation can be extended as follows:

\small
$$ 
\begin{array}{ll}
\textbf{Operation} 						&	RouteReserving \;(\textit{t}:\texttt{T}, \textit{r}:\texttt{R}) \\
\textbf{Rely} 					  		& availability(r) = available \\
															& \forall \; b \; \in \; blocks(r) \; (status(b)=free)  \\
\textbf{Rely} \approx   			& availability(r) = available \\
\textbf{Guarantee}  					& availability(r):=reserved  \\ 
															& \forall \; b \in \; blocks(r) \; (status(b):=occupied) \\
															& route(t):=r \\
															& \forall \; p \in \; P \; (direction(p):=orientation(r)(p)) \\
\textbf{Guarantee} \approx 		& availability(r):= maintenance!  \\ 
															& \forall \; b \in \; blocks(r) \; (status(b):=occupied) \\
															& route(t):=null \\									
\end{array}
$$
\normalsize
This specification includes the case in which, although the requested route is available, not all the related blocks have been freed (for example in one block a sensor stopped working). This is a warning situation and the route needs to be put under observation, the train will be assigned to a null route and, for safety reasons, all the blocks in that route will be occupied. An additional layer of R/G has been added for this purpose and it has been indicated by $\approx$. 

\subsection*{The "make-it robust" process}

The process of adding further layers to the specification considering situations that are abnormal (in the sense that they happen less frequently) is called "make-it robust" process and it will be fully developed and formalized as future work. It is out of the scope of this paper to explain in detail the formalism behind it, this work represents just an introduction to the method with an explanation of the need for it and its potential application to dependable systems. Anyway, the idea we are working on is to modify the global state, passing from what we call the Ideal World (the initial layer) to what we call the Real World (the further layers, it will never be "real" anyway) according to specific formal rules that have to be applied. In this way we restrict the creative act behind the addition of new layers but we make it possible to automatize the consistency check between different layers. Looking at the Polya's analysis of ancient Greeks problem solving \cite{Polya71}, he divides mathematical problems into two classes: "problems to prove" and "problems to find". We have been inspired by this analysis when working on this process. The idea is simply applied: the creative act of identifying the next layer is a "problem to find" and it needs human intervention and invention. This is the hard part of the work. This process is formally guided by a number of rules explaining how the global state, its mappings, the relative domains and ranges and the R/G conditions have to be modified giving a significant spectrum of possibilities, but not infinite freedom. The easy part of the work will be then performed automatically and it will be the "prove" part, the consistency check which represent the automatic correctness analysis.       

\section{Achievements}
In this paper we worked toward an improvement of the ideas presented in \cite{JonesHJ07}. The main contributions can be considered:\\

\begin{enumerate}
\item An understanding of what a method is and an analysis of the desiderata
\item A formalization of the method in \cite{JonesHJ07} and of the features it has to exhibit 
\item EI as a model of faults (and consequent introduction of fault tolerant behavior)
\item The organization of the specification in terms of layers of RG conditions (LFTS)
\item The experimentation on a practical case study
\end{enumerate}

\section*{Acknowledgments} This work has made been possible by the useful conversations with Cliff Jones, Michael Jackson, Ian Hayes, Ani Bhattacharyya, Alexander Romanovsky, John Fitzgerald, Jeremy Bryans, Fernando Dotti, Alexei Iliasov, Ilya Lopatkin and Rainer Gmehlich and it has been funded by the EU FP7 DEPLOY Project \cite{DEPLOY}.

\small

\bibliographystyle{plain}	
\bibliography{refs}

\begin{thebibliography}{10}

\bibitem{DEPLOY}
Deploy: Industrial deployment of system engineering methods providing high
  dependability and productivity.
\newblock \texttt{http://www.deploy-project.eu/}.

\bibitem{AbrialB-book}
J.-R. Abrial.
\newblock {\em The B-book: assigning programs to meanings}.
\newblock Cambridge University Press, New York, NY, USA, 1996.

\bibitem{AbrialEvent-B}
J.-R. Abrial.
\newblock {\em Modeling in Event-B: System and Software Engineering}.
\newblock To be published in 2009.

\bibitem{AbrialZ-book}
J.-R. Abrial, S.A. Schuman, and B.~Meyer.
\newblock {\em A Specification Language}.
\newblock Cambridge University Press, New York, NY, USA, 1980.

\bibitem{BaetenPA}
J.~C.~M. Baeten.
\newblock A brief history of process algebra.
\newblock {\em Theor. Comput. Sci.}, 335(2-3):131--146, 2005.

\bibitem{BjornerVDM78}
D.~Bjorner and C.B. Jones, editors.
\newblock {\em The Vienna Development Method: The Meta-Language}, volume~61 of
  {\em Lecture Notes in Computer Science}. Springer, 1978.

\bibitem{BoochOO}
G.~Booch.
\newblock {\em Object-Oriented Analysis and Design with Applications (3rd
  Edition)}.
\newblock Addison Wesley Longman Publishing Co., Inc., Redwood City, CA, USA,
  2004.

\bibitem{Jones05t}
J.W. Coleman and C.B. Jones.
\newblock Examples of how to determine the specifications of control systems.
\newblock In A.~Romanovsky M.~Butler, C.~Jones and E.~Troubitsyna, editors,
  {\em Proceedings of the Workshop on Rigorous Engineering of Fault-Tolerant
  Systems ({REFT} 2005)}, 2005.

\bibitem{ColletteJ00}
P.~Collette and C.B. Jones.
\newblock Enhancing the tractability of rely/guarantee specifications in the
  development of interfering operations.
\newblock In {\em Proof, Language, and Interaction}, pages 277--308, 2000.

\bibitem{DeFleur93}
M.L. DeFleur, P.~Kearney, and T.G. Plax.
\newblock Mastering communication in contemporary america.
\newblock 1993.

\bibitem{FowlerUML}
M.~Fowler.
\newblock {\em UML Distilled: A Brief Guide to the Standard Object Modeling
  Language, Third Edition}.
\newblock Addison-Wesley Professional, 2003.

\bibitem{JacksonPF}
M.~Jackson.
\newblock {\em Problem frames: analyzing and structuring software development
  problems}.
\newblock Addison-Wesley Longman Publishing Co., Inc., Boston, MA, USA, 2001.

\bibitem{Jones83a}
C.B. Jones.
\newblock Tentative steps toward a development method for interfering programs.
\newblock {\em ACM Trans. Program. Lang. Syst.}, 5(4):596--619, 1983.

\bibitem{JonesHJ07}
C.B Jones, I.J Hayes, and M.A. Jackson.
\newblock Deriving specifications for systems that are connected to the
  physical world.
\newblock In {\em Formal Methods and Hybrid Real-Time Systems}, pages 364--390,
  2007.

\bibitem{AndersonFT}
P.A. Lee and T.~Anderson.
\newblock {\em Fault Tolerance: Principles and Practice}.
\newblock Springer-Verlag New York, Inc., Secaucus, NJ, USA, 1990.

\bibitem{SMART}
Mannion M. and Keepence B.
\newblock Smart requirements.
\newblock {\em SIGSOFT Softw. Eng. Notes}, 1995.

\bibitem{Polya71}
G.~Polya.
\newblock {\em How to Solve It}.
\newblock Princeton University Press, 1971.

\bibitem{Descartes}
L.J. Lafleur~(trans.) R.~Descartes.
\newblock {\em Discourse on Method and Meditations}.
\newblock New York: The Liberal Arts Press, 1960.

\bibitem{SzyperskiCS}
C.~Szyperski.
\newblock {\em Component Software: Beyond Object-Oriented Programming}.
\newblock {Addison-Wesley Professional}, 1997.

\end{thebibliography}

\end{document}